# Resistive and ballistic phonon transport in ß-Ga$_2$O$_3$

R. Ahrling[1], R. Mitdank[1], A. Popp[3], J. Rehm[3], A. Akhtar[3], Z. Galazka[3] and S. F. Fischer[1, 2]

[1] *Novel Materials Group, Humboldt-Universität zu Berlin, Newtonstraße 15, 12489 Berlin, Germany*

[2] *Center for the Science of Materials, Humboldt-Universität zu Berlin, Zum Großen Windkanal 2, 12489 Berlin, Germany*

[3] *Leibniz Institut für Kristallzüchtung, Max-Born-Straße 2, 12489 Berlin, Germany*

The anisotropic thermal conductivity and the phonon mean free path (mfp) in monoclinic ß-Ga$_2$O$_3$ single crystals and homoepitaxial films of several µm were determined using the 3$\omega$-method in the temperature range from 10K-300 K. The measured effective thermal conductivity of both, single crystal and homoepitaxial films are in the order of 20 W/(mK) at room temperature, below 30 K it increases with a maximum of 1000 to 2000 W/(mK) and decreases with $T^3$ below 25 K. Analysis of the phonon mfp shows a dominance of phonon-phonon-Umklapp scattering above 80 K, below which the influence of point-defect scattering is observed. Below 30 K the phonon mfp increases until it is limited by the total ß-Ga$_2$O$_3$ sample size. A crossover from resistive to ballistic phonon transport is observed below 20 K and boundary effects of the total sample size become dominant. This reveals that the homoepitaxial film-substrate interface is highly phonon-transparent. The resistive and ballistic phonon transport regimes in ß-Ga$_2$O$_3$ are discussed corresponding to the models of Callaway and Majumdar, respectively.

## I. Introduction

Gallium oxide (Ga$_2$O$_3$) is a transparent ultra-wide bandgap (4,7-4,9 eV) [10-12] semiconductor of topical research interest for deep UV-devices, gas sensors and high power electronic applications [1-9] with a predicted breakthrough electric field of $E_b$ =8MV/cm, [5] exceeding that of currently used materials in high power electronics ($E_b \approx$ 2,5 MV/cm for SiC and $E_b \approx$ 3,3 MV/cm for GaN) [13]. The most stable form is the polymorph. ß-Ga$_2$O$_3$.

However, a major challenge in electronic device design is heat dissipation due to the low room temperature thermal conductivity which is approximately a factor of 8 and 30 lower than those of bulk GaN and SiC, respectively [14]. An anisotropic thermal conductivity of single crystalline ß-Ga$_2$O$_3$ due to the monoclinic crystal structure has been reported at room temperature: The highest value is determined along the [010] direction of 27 W/(mK) via time domain thermo reflection [15], 22 W/(mK) via laser-flash [16] and 29 W/(mK) with 2$\omega$- and 3$\omega$- measurements [17,18]. Below room temperature, thermal conductivities of 200-300 W/(mK) at 80 K [15] and 190 W/(mK) at 60 K [17] have been reported. The low temperature regime (below 60 K) which is relevant for point-defect and boundary-limited heat transport remains an open issue.

Here, we investigate the thermal transport properties in the full temperature regime between 10 K and 300 K of single crystalline substrates and homoepitaxially-grown thin films as typically used in high-power electronic devices. We identify the scattering mechanisms and the cross-over from resistive to ballistic phonon transport

In general, *heteroepitaxially* grown ß-Ga$_2$O$_3$ films exhibit a reduced room temperature thermal conductivity with decreasing film thickness compared to bulk. For a 2µm film, a reduction of 50% [19], and for 400 nm thin ß-Ga$_2$O$_3$ exfoliated films on quartz glass a reduction of 35% have been observed [20]. In order to maintain the maximal thermal conductivities, single crystals and homoepitaxially grown films of high crystalline quality and purity are required, since defects lower the thermal conductivity [17,18,21]. However, thermal transport through epitaxial interfaces realized by *homoepitaxially* grown ß-Ga$_2$O$_3$ films on a ß-Ga$_2$O$_3$ single crystal substrate has not yet been reported.

While intrinsic phonon scattering consisting of Umklapp- and point defect scattering is known to dominate the thermal transport for temperatures above 100 K [17,18], a crossover from resistive to ballistic phonon transport at lower temperatures may occur in high-quality ß-Ga$_2$O$_3$-single crystals, and as we demonstrate in this study also in homoepitaxially grown epi-layers on single crystalline substrates. In the ballistic regime, phonons do not undergo resistive scattering and the two crystal surfaces are in a thermal equilibrium emitting and absorbing phonons [22]. This is described as phonon-radiative transport (PRT) [23,24]. For this case, the sound velocity is the only relevant material parameter, while the specific heat of the crystal no longer plays a role. The properties of the upper and lower surfaces fully determine the heat transport [24]. The transport regimes are expressed by different relevant mean free paths (mfp) as introduced below. At higher temperatures, the phonon transport is resistive. In the Callaway model [25] a mean free path $\Lambda_C$ is defined, calculated with Matthiessens rule as

$$\frac{1}{\Lambda_C} = \frac{1}{\Lambda_U} + \frac{1}{\Lambda_{PD}} + \frac{1}{\Lambda_B}, \tag{1}$$

where $\Lambda_U$ depicts the mfp by sole contribution of Umklapp scattering, $\Lambda_{PD}$ depicts the mfp by a sole contribution of point-defect scattering and $\Lambda_B$ is the mfp attributed to the contribution of boundary scattering.

The intrinsic mean free path $\Lambda$ is defined for a theoretically infinite crystal, without the influence of any boundary effects. Only Umklapp and point defect scattering are considered $\frac{1}{\Lambda} = \frac{1}{\Lambda_U} + \frac{1}{\Lambda_{PD}}$. If the intrinsic mean free path compares with the sample thickness, the heat transport from heater to the backside of the crystal is ballistic. A model well describing the transition from resistive to ballistic transport is given by Majumdar [24]. Here, the resistive scattering model from Callaway [25] merges with the ballistic transport model from Casimir [22]. An effective mean free path

$$\Lambda_{\text{eff}} = \frac{\Lambda}{1+\frac{4}{3}\cdot\frac{\Lambda}{d}} \tag{2}$$

is introduced by Majumdar [24].

The phonon-radiative-transport as described in the model of Majumdar is formally exchanged by an adequate boundary scattering model. For $\Lambda = d$ the effective mean free path then becomes $\Lambda_{\text{eff}} = \frac{3}{7}d$. In the case of $\Lambda \gg d$, $\Lambda_{\text{eff}}$ approximates instead to $\frac{3}{4}d$, which is often named Casimir limit [22].

For both, the resistive transport regime, as well as the ballistic transport regime, a $T^3$-dependence of the thermal conductivity is expected [23,24], however, for different physical reasons. The principal difference is the following: The thermal conductivity in the resistive regime has its temperature dependence due to the specific heat as $C \propto$ $T^3$. In the regime of ballistic phonon transport as described by Majumdar [24], a thermal equilibrium of emission and absorption of phonons between heater and substrate is established and the $T^3$-dependence of the thermal conductivity is due to the radiation laws which constitute the regime of PRT [22]. Only the sound velocity remains as relevant material parameter [24]. Therefore, the $T^3$ relation is necessary, but not sufficient for the identification of ballistic phonon transport. However, if additionally, the determined mean free path compares with the sample thickness, ballistic transport takes place.

To tackle the problem of heat dissipation in ß-Ga$_2$O$_3$ devices, further research about the heat flow through ß-Ga$_2$O$_3$ interfaces is necessary [14]. For sufficiently high-quality material, the heat transport is dominated by Umklapp scattering for temperatures above 100 K [17,18]. For lower temperatures (depending on the sample thickness), heat can be transported ballistically through the crystal. In this case, the properties of surface and substrate determine the energy transport [22]. For this reason, we expand the previously reported lower temperature limit for thermal conductivity measurements on ß-Ga$_2$O$_3$ from 60 K [17] to 10 K to observe ballistic effects and draw conclusions about the interactions between phonons and interfaces.

In this work, the thermal conductivity and phonon mean free path in ß-Ga$_2$O$_3$ single crystals and homoepitaxial ß-Ga$_2$O$_3$ thin films have been investigated. $3\omega$ measurements have been performed between 10 K and 300 K. Based on these data, the resistive and ballistic phonon transport regimes are identified.

## II. Experiment

Crystal samples for the present study were prepared from Czochralski grown crystals as described in detail elsewhere [26-29]. Electrically insulating crystals by doping with Mg were oriented, cut into 10×10 mm$^2$ bars in the cross-section, and cleaved parallel to the (100) plane.

For the growth of thin homoepitaxial films, the (100) oriented substrates have been polished with a 4° off cut prepared toward the [00-1] direction to which provides vicinal surfaces to allow for a layer growth in smooth step-flow growth mode. In the next step, homoepitaxial (HE) films of 2 µm and 3 µm ß-Ga$_2$O$_3$ have been grown on the substrates via MOVPE growth. [30,31] These homoepitaxial layers have been compared with a bulk crystal (SCB) with the scope of this work.

XRD measurements show, that the highest crystal lattice quality is achieved for the single crystalline bulk substrate sample (SCB) with a full width at half maximum (FWHM) of 29 arcsec, as shown in figure 1a. This is in good agreement with previously reported MOVPE samples of high growth quality [31]. The structural quality of the HE films has been measured to either be of similar quality (FWHM 31 arcsec for the SCB+2µmHE sample) or only slightly decreased compared to the bulk (FWHM 45 arcsec for the SCB+3µmHE sample). The morphology of the grown layer has been measured by AFM. A representative picture is shown in Fig. 1b for a 3µm thick HE grown ß-Ga$_2$O$_3$ layer. It can clearly be seen that in this case the desired step flow growth mode has not been stabilized up to 3 µm. As a result a superimposing of several steps occurred leading to the here shown step bunching effect.

The roughness of the polished sample surfaces has been determined by AFM to be in the order of 1 nm while the roughness of the unpolished backside is in the order of 5 µm. TEM imaging (see e.g. Fig. S4, supplementary material) and comparison to other samples revealed no visible accumulation of defects at the interface between substrate and homoepitaxial layer (for the full details on growth process and crystallographic analysis see [26-30]).

To prepare the measurement structures on the sample surfaces (shown in figure 1c), metal lines were fabricated on top of the samples using photolithography (with positive resist AZ ECI 3027), magnetron sputtering of Ti/Au (5 nm/50 nm) and subsequent lift-off in an ultrasonic bath of DMSO.

In this work, the $3\omega$-method [32,33] was used to determine the thermal conductivity. Running an AC current $I$ with angular frequency $\omega = 2\pi f$ through the outer metal contacts of the heater lines on the sample causes a temperature oscillation $\Delta T$ due to Joule heating. Since the heating power is $P = RI^2$, with $R$ as resistance, temperature oscillations will be $\Delta T \propto \cos(2\omega t)$. Due to the temperature dependence of the metal lines resistance of $R = R_0(1 + \alpha \Delta T)$ the resistance will then also oscillate as $R \propto \cos(2\omega t)$. The temperature coefficient $\alpha$ and the resistance $R_0$ are functions of the bath temperature $T_0$, as seen in figure S1. The voltage between the inner contacts $U = RI$ then has two components, one $U_{1\omega} \propto \cos(\omega t)$ and one $U_{3\omega} \propto \cos(3\omega t)$. The $3\omega$-voltage signal contains information about the temperature increase $\Delta T$ and thermal conductivity $\lambda$ of the sample. Assuming $\alpha \Delta T \ll 2$ (this condition is proven to hold true over the entire temperature range, see figure S1), the temperature oscillation can be experimentally determined to be

$$\Delta T = \frac{2 U_{3\omega}}{\alpha U_{1\omega}}. \tag{3}$$

It can be shown, that

$$\Delta T = \frac{P}{l\pi\lambda}\left(\frac{1}{2}\ln\frac{D}{b^2} + \ln 2 - 0{,}5772 - \frac{i\pi}{4} - \frac{1}{2}\ln 2\omega\right) \tag{4}$$

where $D$ is the thermal diffusivity and $b$ the heater width [32,34].

Combining these equations, the thermal conductivity can be extracted from the slope of $U_{3\omega}$ vs. $\ln 2\omega$ and it is $\lambda = \frac{\alpha U_{1\omega}^3}{4\pi l R} \cdot \frac{\Delta \ln 2\omega}{\Delta U_{3\omega}}$. The heating voltage is provided by the output of a SR 830 lock-in amplifier. With two such lock-in amplifiers, the $1\omega$- and $3\omega$-voltage signals are measured simultaneously. The sample heater line is put in series with a potentiometer, which is set to the same resistance as the sample to filter out the $1\omega$ part by using a differential amplifier. This is done to be able to detect the much smaller $3\omega$-signal. A depiction of the measurement setup is seen in figure 1c.

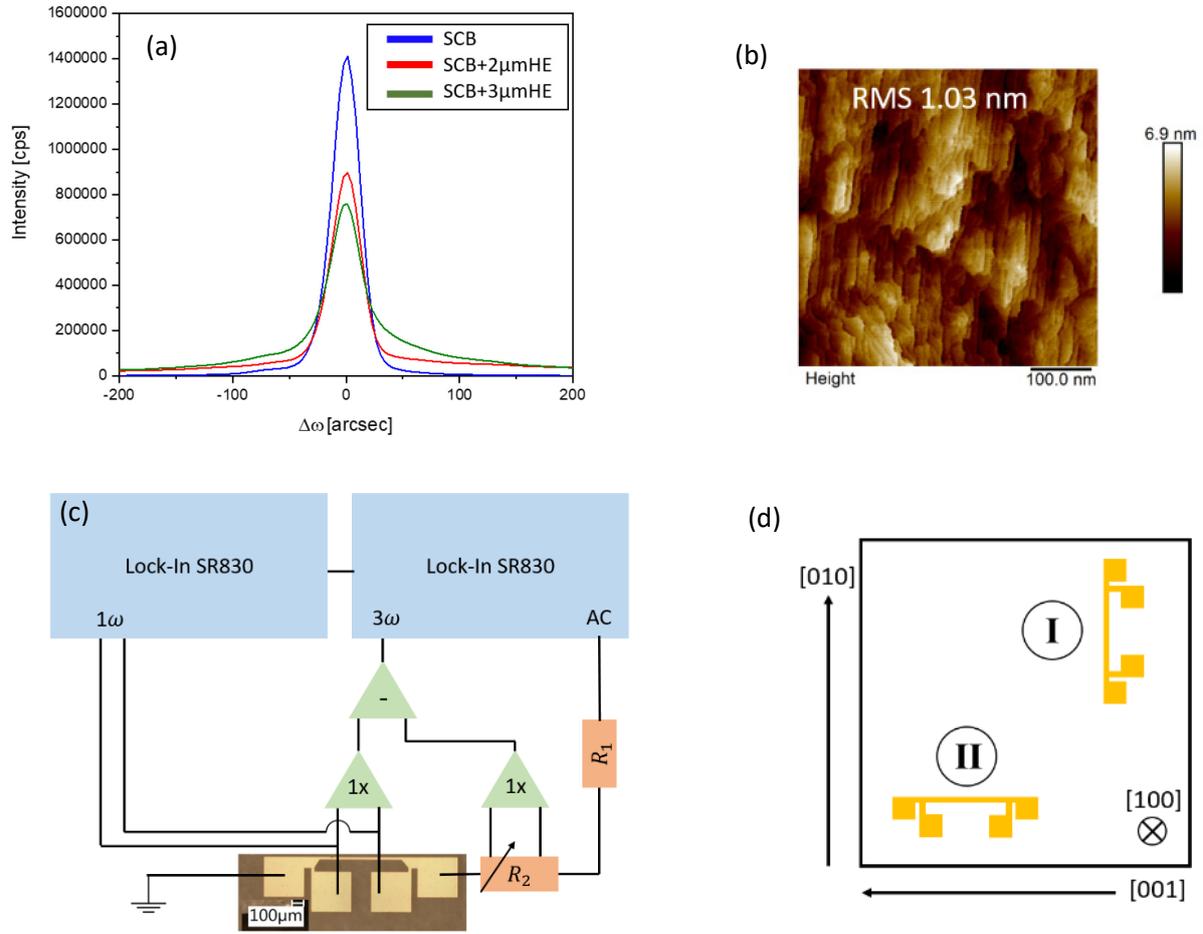

FIG. 1. Overview of the material quality, measurement setup and sample geometry. (a) XRD measurement of the single crystalline bulk (SCB) sample (FWHM 29 arcsec), SCB+2μmHE sample (FWHM 31 arcsec) and SCB+3μm sample (FWHM 45 arcsec) (b) AFM image of the sample surface morphology after homoepitaxial growth of a 3μm film. **(c)** Measurement setup for the 3ω-voltage measurements. (d) Schematic of the measurement structures on the sample, measuring in two different geometries: **I** denotes measurements of the heat flow in the combined [100]/[001] direction, **II** denotes measurements of the heat flow in the combined [100]/[010] direction.

Temperature-dependent measurements are performed in a KONTI IT flow cryostat. To minimize influences of thermal convection of the surrounding He-atmosphere, the sample chamber is held under vacuum. The influence of thermal radiation can be neglected in the temperature range examined in this work [35].

3ω-voltage measurements were performed on each of the 3 samples, one insulating single crystalline bulk sample (SCB) and two samples with a conductive homoepitaxial layer on top (SCB+HE). Each sample had two heater orientations, see figure 1c. For thermally anisotropic samples, the 3ω-voltage measurement detects the mean thermal conductivity of the two directions perpendicular to the heater (in plane and cross plane).

### III. Results

Representative 3ω-voltage measurements are shown in figure 2. The linear dependence of 3ω-voltage and logarithmic frequency holds up for all temperatures in their corresponding frequency bands. These measurements confirm the linear relation between $U_{3\omega}$ and $\ln 2\omega$. It can be seen, that the gradient $\Delta U_{3\omega}/\Delta \ln 2\omega$ decreases for lower temperatures, which indicates a rise in thermal conductivity. Below 30 K this gradient increases, indicating a decline in thermal conductivity. These temperature dependencies will be discussed in the following part. For the low temperature plot, the low frequency data show a different slope corresponding to the sample holder, yielding a similar value for all shown temperatures. Here, the thermal penetration depth exceeds the sample thickness.

For room temperature (here 20°C, 293 K) the measured effective thermal conductivities have been determined as $\lambda_{[100]/[010]} = (20 \pm 2)$ W/(mK), with good agreement between all measured samples and as $\lambda_{[100]/[001]} = (19 \pm 2)$ W/(mK). Overall, this confirms previous reports [14,15,17,37] in which the thermal conductivity is highest in the [010] direction under the assumption that the thermal conductivity in the [100] is the lowest.

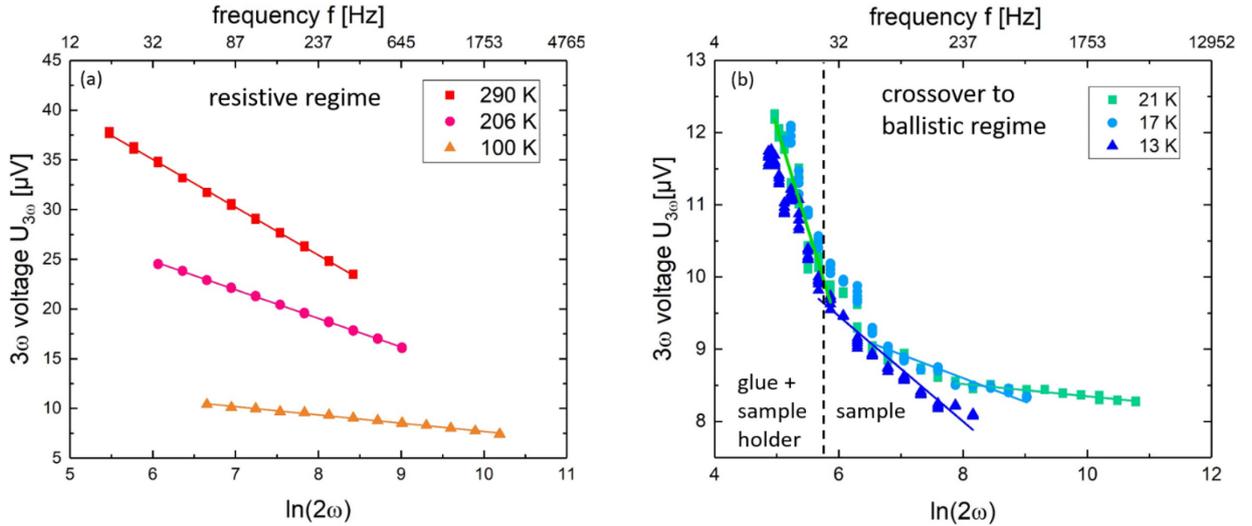

FIG. 2. Measured 3ω Voltage as a function of the logarithmic frequency. The data are taken from the 3μm epitaxial film sample in the [100]/[010] configuration (II). (a) Resistive thermal transport regime (Umklapp) for high temperatures (290 K – 100 K) (b) Crossover to the ballistic thermal transport regime at low temperatures (21 K – 13 K). In both cases, the linear dependence $U_{3\omega} \propto \ln(2\omega)$ can clearly be seen. Different frequency bands used for evaluation stem from a temperature-dependent thermal penetration depth.

However, in this case the SCB+3μm HE sample showed a reduced value of $\lambda_{[100]/[001]} = (14 \pm 1)$ W/(mK). This reduced value for $\lambda_{[100]/[001]}$ in the 3 μm sample may occur due to growth defects in the epitaxial layer induced by the step bunched growth morphology (see Fig 1b). These defects tend to elongate in the (001) planes which is expected since they are macro steps resulting from destabilized step flow growth. The step bunched surface morphology could explain a reduction in heat transport in dependence of the crystal direction [36]. The deviation occurs only above 200 K. Here, the phonon mfp is much smaller than the epitaxial layer´s thickness and the thermal penetration depth is of comparable order of magnitude (see supplementary material S7). Therefore the 3ω-voltage measurement is sensitive to the defects in the layer in this temperature regime.

To compare the thermal conductivity results with previously reported literature values it has to be considered that a measurement in 3ω-geometry on an anisotropic crystal will always yield the mean value of both directions perpendicular to the heater $\lambda_{xy} = \frac{\lambda_x + \lambda_y}{2}$. Values calculated from literature in this way are compared with the results of this study and can be seen in the table S2.1 in the supplementary material. For configuration II, the measurements are well in accordance with previously reported values. For configuration I (see supplementary material S5), the measurements exceed our previously measured values [14,15,17,18] by roughly 10 %, while the theoretically predicted value [37] is in agreement. This indicates a higher material quality in this study, as explained later.

As expected, the thermal conductivity increases with decreasing temperature, see figure 3. Below 80-90K a change in the slope of the double logarithmic plot can be observed, indicating a change in the dominant phonon scattering mechanism. Here, the anisotropy between both measurement configurations also diminishes. A maximum thermal conductivity is reached for all samples around 25-30 K with values of 1000-3000 W/(mK).

Below that, the thermal conductivity decreases with decreasing temperature. In this regime, a $T^3$-dependence of the thermal conductivity is expected as detailed in the introduction.

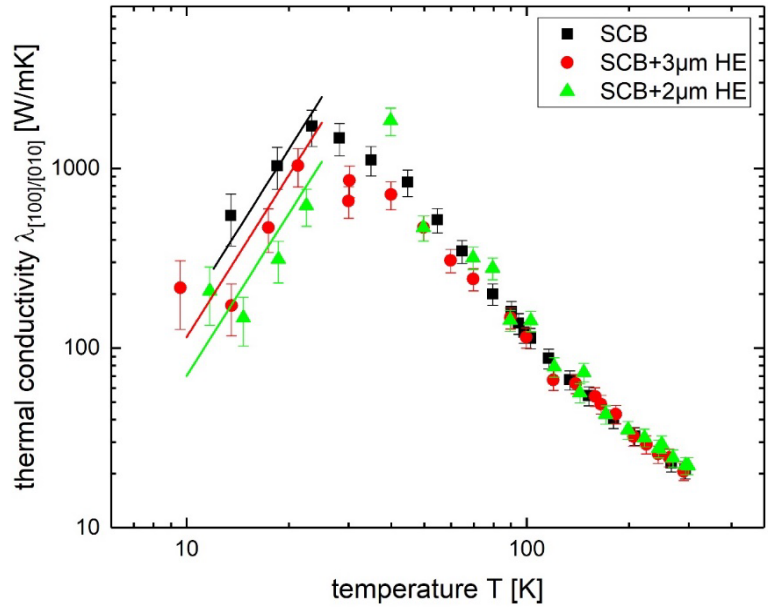

FIG. 3. Measured thermal conductivity in the [100]/[010] configuration II as a function of the temperature. Data of three samples are shown: SCB (black squares), SCB+3μm HE (red circles) and SCB+2μm HE (green triangles). The lines act as a guide for the eye.

## IV. Discussion

From the measured thermal conductivities, the effective phonon mfp can be calculated as

$$\Lambda_{\text{eff}} = \frac{3\lambda}{v_S c_p \rho} \tag{5}$$

Since the measured thermal conductivities are a mean of two directions, the values of the sound velocities (Guo *et al.* [14]) are applied in the same way.

The overall temperature dependence of the phonon mean free path can be seen in figure 5. It increases with a decrease in temperature. Below 25-30 K, the mean free path begins to flatten out, reaching a plateau at a value of 3/7d = 214 µm, indicating a crossover from resistive to ballistic phonon transport.

A fit for the mean free path over the entire temperature range was applied by a combination of phonon-phonon-Umklapp scattering and point-defect scattering (as given by the Callaway-model [25]) for the intrinsic scattering:

$$\frac{1}{\Lambda} = \frac{1}{\Lambda_U} + \frac{1}{\Lambda_{PD}}$$

The Umklapp scattering is given by

$$\Lambda_U \propto \left(e^{\Theta_D/2T} - 1\right). \tag{6}$$

Between 30 K and 80 K, point defect scattering plays a relevant role and it is modelled as

$$\Lambda_{PD} \propto T^{-4}, \tag{7}$$

analogous to Rayleigh scattering.

Below 30 K, scattering at the surface boundaries becomes dominant and if the effective mean free path $\Lambda$ compares with the overall sample thickness $d$, a constant mean free path of $\Lambda_B = 3/7d$ according to (1) is used to describe the low temperature limit in the formalism of the Callaway model (2). This is discussed in detail below. All curves fitted with the Callaway model are in good agreement with the measured data.

For temperatures above 80 K, Umklapp scattering dominates the phonon transport, as shown in figure 4. It can be seen, that contrary to the thermal conductivities, there is no anisotropy of the mean free path, as expected. An Umklapp scattering fit was applied to the high temperature data, modelling the points well above 80 K. This means, no systematic influence of defects is observed here and the obtained mean free paths have the quality of material parameters. The phonon mfp for that regime can be expressed by

$$\Lambda = A \cdot \left(e^{\Theta_D/2T} - 1\right) \tag{8}$$

with the parameter $A$ representing the value of $\Lambda(\Theta_D/2)$, with $A \approx (0{,}8 \pm 0{,}1)$ nm. A least-square fit of the Umklapp scattering gives a Debye temperature of $\Theta_D = (975 \pm 50)$ K. Previously a value of $\Theta_D = 870$ K was predicted [38].

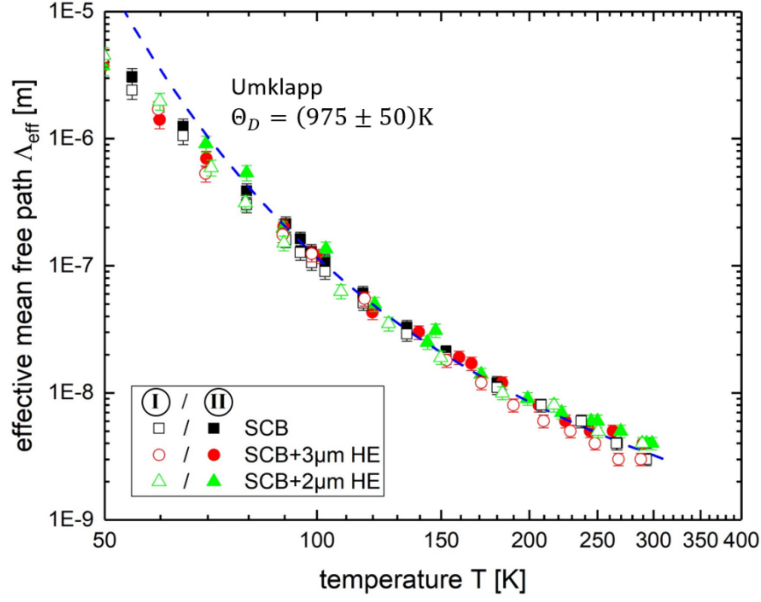

FIG. 4. Effective mean free path as a function of temperature for all measured samples. Both configurations I and II are shown. The effective phonon mean free paths show no anisotropy. The model of Umklapp scattering (blue dashed line) was used to describe the temperature regime above 80 K. Deviations from this model begin only below 80 K.

Below 80 K point-defect scattering dominates the transport. Previously, we showed that point defects already play a role at 150 K [16] which indicates a higher purity of the material investigated in this study. In the regime of point defect scattering, the anisotropy of the thermal conductivity vanishes as expected. The fit for the point defect scattering is calculated as

$$\Lambda_{\text{PD}} = \Lambda(T_0)\left(\frac{T_0}{T}\right)^4. \qquad (9)$$

The measurement data show a good agreement with the $T^{-4}$-Rayleigh model between 30 K and 80 K. This fit can be compared with the relation $\Lambda_{\text{PD}} = \frac{1}{\Phi n}$ [24], with $n$ as density of point defects and $\Phi$ as scattering cross section. For low temperatures, the scattering cross section can be expressed as $\Phi = \frac{16\pi^5 R^6}{l_{dom,0}^4 T_0^4} T^4$ [24], with $R$ as atom radius of the scattering centers and $l_{dom,0}$ as dominant phonon wavelength (see figure S3, supplementary material) at temperature $T_0$. With these relations, the density of scattering centers in the material can be approximated as

$$n = \frac{l_{dom,0}^4}{16\pi^5 R^6 \Lambda(T_0)} \qquad (10)$$

using the fit parameter $\Lambda(T_0)$ with the value of $\Lambda(50\text{ K}) \approx 6{,}4$ µm. A comparison between the known atomic radii [41] and the expected concentrations of each potential atom, the most likely candidate acting as a scattering center are the $^{18}$O isotopes within the ß-Ga$_2$O$_3$ crystal (see S6, supplementary material).

The resistive (intrinsic) scattering regime above 30 K can be well described by the combination of Umklapp-scattering and point-defect-scattering in the Callaway model [25].

In detail, we discuss the low-temperature observations below. At low T<30 K the effective phonon mfp approximates to a constant value. The fits in the low temperature regime are in good agreement with the data, as shown in figure 5.

To describe the low temperature limit in the formalism of the Callaway model [25], $\Lambda_B$ is substituted by 3/7 $d$ [17] (see equation (2)). The experimental data are well described by this fixed $\Lambda_{\text{eff}}$ at low temperatures, as can be seen in figure 5. This requires an *intrinsic* phonon mfp in the samples $\Lambda >\approx d$ comparable to the film thickness. The occurrence of both observations, first, the phonon mfp reaching a plateau at a value of $\frac{3}{7}d$ and, second, the $T^3$ dependency in this temperature range, indicates that the transition from resistive to ballistic phonon transport is observed. Therefore, diffusive interactions with the crystal´s surface of the back side take place, leading to phonon-*radiative* heat transport across the crystal. In such a case, resistive scattering processes are negligible and the interaction with the crystal surfaces dominates the thermal transport properties. However, a phonon mfp of $\frac{3}{4}d$, which would correspond to the case of $\Lambda \gg d$ in the Casimir limit is not quite achieved, as can be seen in figure 5. Nevertheless, it is approached up to 60 % at 10 K ($\Lambda >\approx d$).

For the SCB+2µm HE sample the effective phonon mfp approximates to 60 µm which is much less than the crystal thickness. This can be explained by the existence of defects in the crystal which limit the phonon mfp. The temperature dependence of the effective phonon mfp below 30K is determined by the specific heat so that $\Lambda_{\text{eff}}(T) \propto T^3$.

Using the modeled phonon mfps, the thermal conductivity data can be fitted, using the relation $\lambda = \frac{1}{3}\Lambda c_P \rho v_S$. This has been performed for all samples as shown in figure 5b, 5d and 5f. The fits represent the measured thermal conductivity data well for the entire temperature range.

In the context of ballistic phonon transport the interaction between the phonons and the crystal surface is of high importance. Therefore, the surface properties are discussed here. The backsides of the samples (substrates) were not polished, leading to a surface roughness of several µm determined by atomic force microscopy. To avoid the occurrence of specular reflection at the backside in the case of ballistic phonon transport, the dominant phonon wavelength $l_{\text{dom}}$ must be smaller or comparable the roughness $r$ to ensure diffuse emission. The dominant phonon wavelength can be calculated by a modified version of Wiens law to be $l_{dom} = hv_s/(2{,}82\, k_B T)$ with the sound velocity $v_s$ ($v_s = 4318\, m/s$ in [100] direction) and temperature $T$. For $T = 20$ K, this gives a value of $l_{dom} \approx$ 4nm. This is illustrated in figure S2. Since clearly $l_{dom} \ll r$, the interaction of phonons with the surface is expected to be diffusive. The surface of the epitaxial layers has a considerably lower roughness (see figure 1). In this case $l_{dom} \approx r$. Furthermore, as required it is ensured, that the temperature on the upper surface of the sample is the heater temperature as it is in direct contact with the heater line. According to Majumdar [24], the electron-phonon mean free path in the metal is much smaller, than the phonon mfp in the film and thus the interface of the dielectric film can be assumed to be a thermalizing black boundary.

As required, on the crystal backside the contact is to the bath temperature. Here, a different situation is prevalent: Due to the roughness r $<\approx$ 5 µm of the sample backside the diffusive mismatch model can be applied [23]. The density of states $N \propto \frac{\omega^2}{v^3}$ (see figure S3, supplementary material) for phonons of a frequency $\omega$ is much higher in the glue below the sample, due to its considerably lower sound velocity. Therefore, practically all phonons are absorbed into the glue and their energy dissipates into the thermal bath. The interface can therefore be assumed to be a thermalizing black boundary fixed at the bath temperature. Thermalized phonons will then be re-emitted diffusively from the rough surface and a radiative equilibrium will be established.

For the SCB and the SCB+3µm HE samples, all conditions of the Casimir model [22] have been proven true below 20 K. The thermal conductivity shows a $T^3$-dependence of the thermal conductivity while the temperature gradient remains small compared to the bath temperature. The surface roughness of the crystals backside is much higher than the phonon wavelength, indicating diffusive interaction between the phonons and the surface. The maximum reached for the effective mean free path fits well to the value of $3/7d$, expected for the case of $\Lambda \gtrapprox d$. We therefore conclude, that the crossover from resistive to ballistic transport is observed. Ballistic phonon transport from the heater to the back surface of the crystal takes place. The heat transport can be described as phonon-radiative-transport, which is no longer dominated by intrinsic material properties but by the crystal surface only.

In the SCB+2µm HE sample, ballistic transport of phonons between the two crystal surfaces was not achieved. However, the effective phonon mfp still greatly exceeded the thickness of the homoepitaxial films. Consequently, the interface between the single crystalline bulk substrate and the MOVPE-grown homoepitaxial layer does not act as a scattering surface for phonons. Heat can be transported through the interface without any noticeable thermal resistance. Therefore, we identify the interface as *phonon-transparent*. It indicates a high epitaxial growth quality at the interface of bulk material and homoepitaxial layer. This is of special interest in the case of electrically conductive layers, where under these assumptions the electrons would be confined to the thin film whereas the phonons can propagate through the entire substrate underneath.

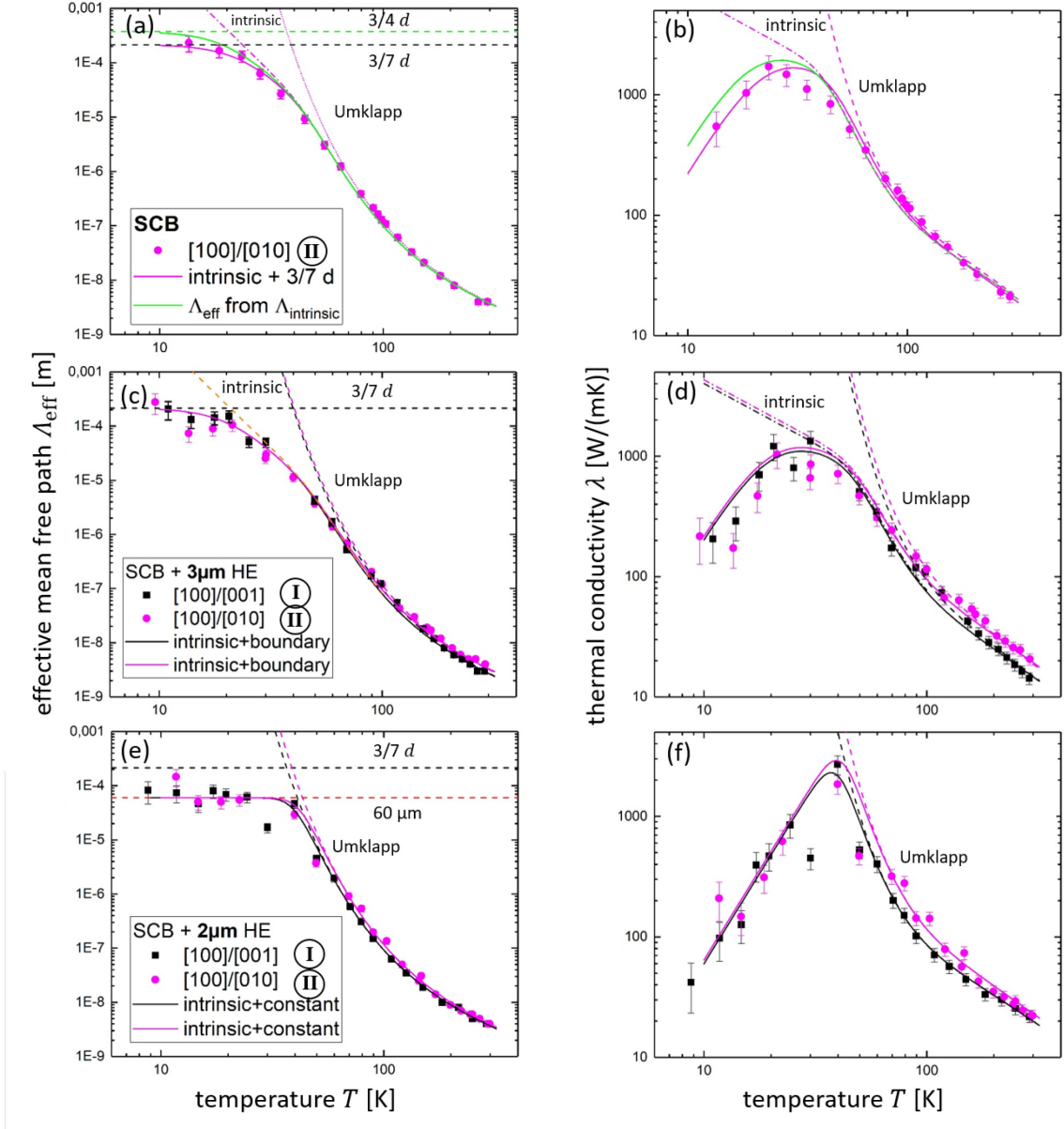

FIG.5. Detailed analysis. Left column (a,c,e): effective mean free paths calculated from measured thermal conductivities fitted after the models of Callaway (high T) and Majumdar (low T). Right column (b,d,f) Fits from the curves in the left column applied to the measured thermal conductivities. (a) Effective Phonon mean free path, calculated from the measured thermal conductivity, as a function of temperature on the SCB sample (configuration II). It can be seen, that a model including Umklapp scattering, point defect scattering and a constant mean free path of $\frac{3}{7}d = 214$ μm fits well the effective mean free path data. (b) Thermal conductivity as a function of temperature on the SCB sample. The data has been modeled by calculating the thermal conductivity from the fit data for the effective mean free path acquired in (a). (c) Effective Phonon mean free path, calculated from the measured thermal conductivity, as a function of temperature. A combination of Umklapp scattering, point defect scattering and boundary scattering has been applied to model the data. The value of $\frac{3}{7}d$ for the boundary scattering models the low temperature data well. (d) Thermal conductivity as a function of the temperature. The data have been modeled by calculating the thermal conductivity from the fit data for the effective mean free path acquired in (c). (e) Effective Phonon mean free path, calculated from the measured thermal conductivity, as a function of temperature. A combination of Umklapp scattering and boundary scattering has been applied to model the data. A reduced value for the maximum of mfp (60 μm < d) was observed, giving rise to the assumption of defects in the substrate limiting the mfp. Due to this reduction, intrinsic scattering

is present in this sample. Nevertheless, the heat transport through the epitaxial layer proves to be ballistically for T< 70 K. (f) Thermal conductivity as a function of the temperature. The data have been modeled by calculating the thermal conductivity from the fit data for the effective mean free path acquired in (e).

The above results show that for sufficient low temperatures heat can be transported ballistically through thin ß-Ga$_2$O$_3$ films. Currently, electronic devices are often designed to use ß-Ga$_2$O$_3$ layer thicknesses in the order of 1 μm. An effective phonon mfp of some μm is reached at temperatures of approximately 70 K, for a high-crystalline quality of ß-Ga$_2$O$_3$ comparable to the samples in this work. At such operating temperatures, near to that of liquid nitrogen (77 K), ß-Ga$_2$O$_3$ has a much higher overall thermal conductivity ($\lambda = (200 \pm 30)$ W/(mK) at 77K) compared to room temperature. This increase in the thermal conductivity is comparable to that of the commonly used high-power material SiC [39]. Advantageously, the electrical conductivity of ß-Ga$_2$O$_3$ at 77 K is only slightly reduced compared to room temperature (factor 2-3). In contrast, SiC shows a drop in the electrical conductivity of an entire order of magnitude [40].

Below 20 K, the observation of ballistic phonon transport along a macroscopic dimension of 500 μm demonstrates the high crystalline and epitaxial quality of the ß-Ga$_2$O$_3$ single crystals *and* homoepitaxial layers leading to a phonon-transparent interface. In summary, our findings show that in high-quality *homoepitaxial* ß-Ga$_2$O$_3$-layers heat transport may be engineered for electronic devices at cryogenic temperatures of liquid nitrogen.

## V.    Conclusion

The room temperature anisotropy of the thermal conductivity of high-quality ß-Ga$_2$O$_3$ single crystals and homoepitaxial films vanishes for temperatures below 80 K, where the role of Umklapp scattering and therefore the monoclinic crystal structure becomes negligible. Between 80 K and 30 K, point defect scattering, presumably by $^{18}$O isotopes, dominates the phonon transport. For temperatures below 20 K boundary effects dominate. The effective phonon mfp has been determined to be in the order of 3 nm at room temperature which, in contrast to the thermal conductivity, was observed to be isotropic. It increases to approximately 1 μm at 80 K and is limited below 20 K by an effective sample size. Modelling of the Umklapp and point-defect scattering succeeds by use of the Callaway model and the size limitation at low temperatures by the Majumdar model. From that we conclude that the interface between substrate and film is phonon-transparent and that we observe the crossover from resistive to ballistic heat transport in the phonon-radiative transport regime.

## VI.    Acknowledgements

This work was performed in the framework of GraFOx, a Leibniz-ScienceCampus, partially funded by the Leibniz association and by the Deutsche Forschungsgemeinschaft (FI932/10-1, FI932/11-1 and PO 2659/1-2). The authors would like to thank Martin Schmidbauer for fruitful scientific discussions.

# Supplementary Material

# Resistive and ballistic phonon transport in ß-Ga$_2$O$_3$

R. Ahrling[1], R. Mitdank[1], A. Popp[3], J. Rehm[3], A. Akhtar[3], Z. Galazka[3] and S. F. Fischer[1,2]

[1] *Novel Materials Group, Humboldt-Universität zu Berlin, Newtonstraße 15, 12489 Berlin, Germany*

[2] *Center for the Science of Materials, Humboldt-Universität zu Berlin, Zum Großen Windkanal 2, 12489 Berlin, Germany*

[3] *Leibniz Institut für Kristallzüchtung, Max-Born-Straße 2, 12489 Berlin, Germany*

**S1) Temperature-dependence of the temperature-coefficient of the resistivity (metal heater lines)**

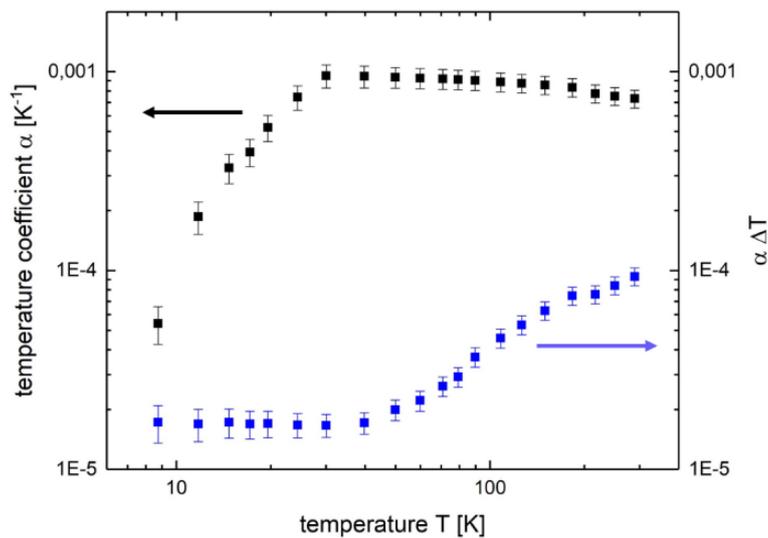

**Figure S1:** Temperature coefficient of the resistivity of the metal lines α and α multiplied with the temperature gradient ΔT as a function of temperature. The temperature coefficient follows the expected behavior for a metal. It can be seen, that $\alpha \Delta T \ll 2$ is fulfilled for every temperature.

## S2) Comparision of effective room temperature thermal conductivities with values from literature

An overview of the effective room temperature thermal conductivities measured in this work can be seen in table S2.1. To compare the results with previously reported literature values it has to be considered that a measurement in 3ω-geometry on an anisotropic crystal will always yield the mean value of both directions perpendicular to the heater $\lambda_{xy} = \frac{\lambda_x + \lambda_y}{2}$. Mean values calculated from literature components $\lambda_x$, $\lambda_y$, $\lambda_z$ in this way are compared with the results of this study in table S2.1. For configuration II, the measurements are well in accordance with previously reported values. For configuration I, the measurements exceed previously measured values. Only the theoretically predicted value is in agreement. Additionally, the experimental value obtained by $2\omega$ and $3\omega$ method only slightly differ, while the values measured with optical methods are both considerably lower.

Summing up, it must be noted, that the variation of the values for the thermal conductivity seem to be due to the crystal quality. Elevated levels of defects in the epitaxial film of the SCB+3µm HE sample have been shown to reduce the thermal conductivity in the Umklapp scattering regime.

TABLE S2.1. ß-$Ga_2O_3$ single crystals: Effective room temperature thermal conductivities measured in this work compared with literature values, both theoretical and experimental. The measurements in configuration II are in good agreement with literature. The measurements in configuration I agree with theoretical predictions. However, they exceed previously reported experimental values.

| Measurement configuration | Thermal conductivity [W/(mK)] (this work) | Thermal conductivity [W/(mK)] (literature) | Reference, method |
|---|---|---|---|
| x/y=[100]/[001] (I) | 18 ± 2 | 16 | [18], 2ω + 3ω |
| | | 13 | [15], TDTR |
| | | 13 | [16], laser flash |
| | | 18 | [39], calculated |
| x/y=[100]/[010] (II) | 21 ± 2 | 20 | [18], 2ω +3ω |
| | | 19 | [15], TDTR |
| | | 18 | [16], laser flash |
| | | 19 | [39], calculated |

## S3) Phonon spectral density as a function of phonon wavelength

The phonon spectral density can be calculated as $N = \frac{\omega^2}{2\pi^2 v_S^3} \cdot \frac{1}{\exp\left(\frac{\hbar\omega}{k_B T}\right) - 1}$

In terms of the phonon wavelength, this translates to $N = \frac{2h}{l^3} \cdot \frac{1}{\exp\left(\frac{hv_S}{lk_B T}\right) - 1}$.

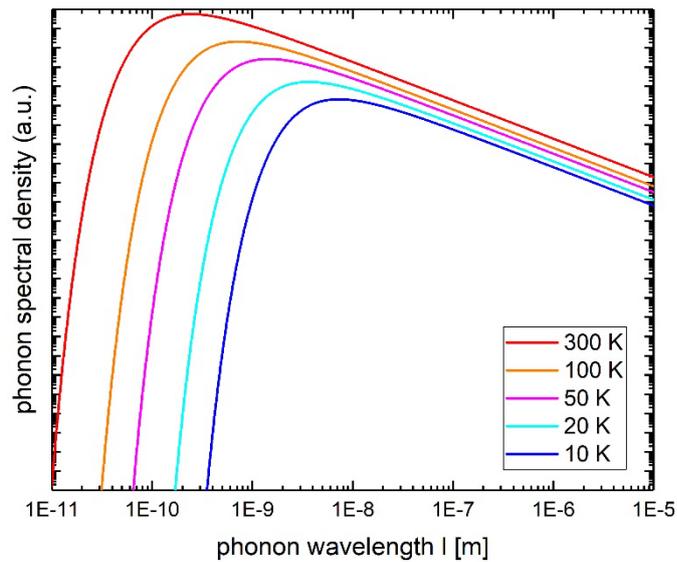

**Figure S3:** Phonon spectral density as a function of phonon wavelength. The wavelength of phonons dominating the heat transport shifts to higher values with decreasing temperature. For all temperatures examined in this work, it is in the range of nanometers and below.

## S4) Transmission electron microscopy (TEM) images

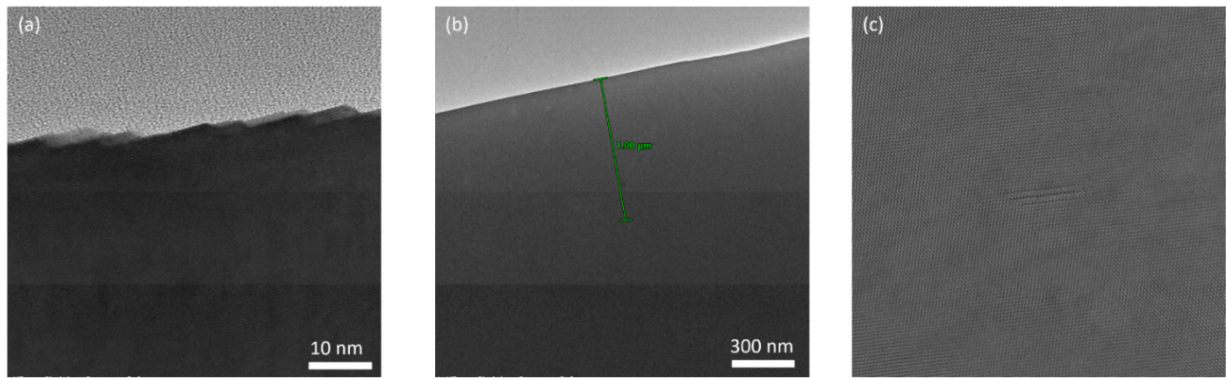

**Figure S4:** TEM images of a ß-Ga$_2$O$_3$ single crystal with a 1μm homoepitaxially grown ß-Ga$_2$O$_3$ layer. (a) TEM image of the surface of the homoepitaxial layer. Well defined step-flow growth is visible indicating a high growth quality. (b) TEM image showing the interface between homoepitaxial layer and substrate. The interface cannot be distinguished in this image. No accumulation of defects could be observed at the interface. (c) For comparison: TEM image of a ß-Ga2O3 sample showing a clearly visible defect. All images were taken at an acceleration voltage of 300 kV.

## S5) Thermal conductivity and phonon mean free path in the [100]/[001] configuration (I)

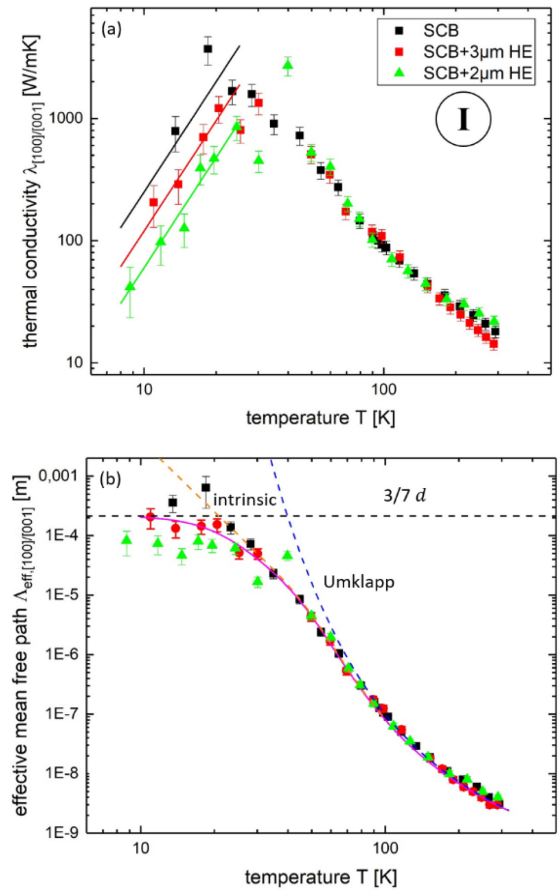

**Figure S5: (a)** Measured thermal conductivity in the [100]/[001] configuration (I) as a function of the temperature. Data of three samples are shown: SCB (black squares), SCB+3µm HE (red circles) and SCB+2µm HE (green triangles). For low temperatures, a $T^3$ function has been applied as expected for ballistic phonon transport. **(b)** Effective Phonon mean free paths in the [100]/[001] configuration **I**, calculated from the measured thermal conductivity, as a function of temperature. A combination of Umklapp scattering and point-defect scattering has been applied to model the intrinsic scattering. The saturation value corresponds to a mean free path of nearly 3/7d as expected for phonon-radiative transport.

## S6) Density of point defects as a function of the atom radius of scattering centers

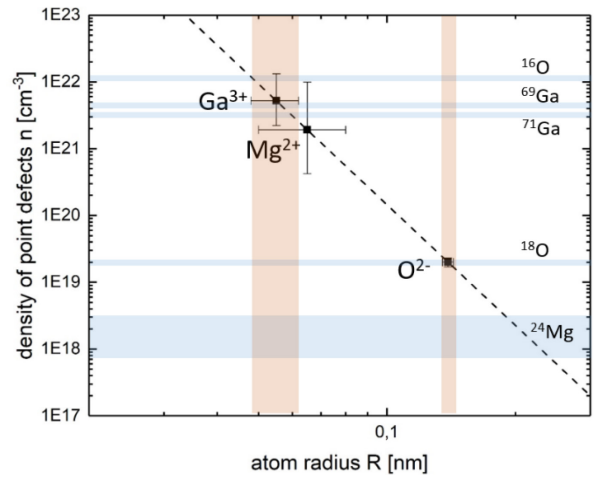

**Figure S6:** Density of point defects as a function of the atom radius of scattering centers. The curve is calculated with formula (10) using $T_0 = 50$ K, $\Lambda(T_o) = 6{,}4$ µm, $l_{dom,0} = 1{,}46$ nm. A comparison between the calculated concentrations (black squares) and the expected concentrations in the crystal (blue bands) and atom radii (red bands) for each material shows an overlap for the $^{18}$O, $^{69}$Ga and $^{71}$Ga isotopes.

The ionic radii in oxide crystals have been taken from ref. [S6.1]. The ionic radius of oxygen is assumed as $R = 0{,}135 - 0{,}142$ nm. The concentration of oxygen atoms in the ß-Ga$_2$O$_3$ crystal can be calculated using the molar mass and density of ß-Ga$_2$O$_3$ to be $1{,}1 \cdot 10^{22}$ cm$^{-3}$. This does not fit the data calculated from the fit model. However, taking into account the $^{18}$O isotope, occurring at a rate of roughly 0,2 %, the model fits well.

Isotopic abundances are taken from [S6.2]. The ionic radius of gallium is assumed as $R = 0{,}047 - 0{,}062$ nm. The concentration of gallium atoms in the ß-Ga$_2$O$_3$ crystal can be calculated using the molar mass and density of ß-Ga$_2$O$_3$ to be $7{,}5 \cdot 10^{21}$ cm$^{-3}$. Gallium is split into two isotopes occurring at similar rates (60% $^{69}$Ga and 40% $^{71}$Ga). These overlap well with the model.

The ionic radius of magnesium is assumed as $R = 0{,}049 - 0{,}082$ nm. The doping concentration in the substrate bulk crystals has been shown to be between $0{,}7 \cdot 10^{18} - 3 \cdot 10^{18}$ cm$^{-3}$. This does not fit the data calculated from the fit model. However, narrowing down the scattering centers to $^{18}$O and $^{71}$Ga (due to its lesser abundance than $^{69}$Ga), their individual contributions to the measured point defect scattering can be estimated. Taking the expected concentration for each isotopes formula (10) can be solved for $\Lambda_0$. The contributions to the overall value of $\Lambda_0 = 6{,}4$ µm are then $\Lambda_{0,O} \approx 6{,}4$ µm and $\Lambda_{0,Ga} \approx 12$ µm. Scattering at the $^{18}$O isotope seems to be the main contribution to the point-defect-scattering process. Combining these two values after Matthiessens rule gives an overall scattering length of $\Lambda_{0,\text{total}} \approx 4{,}2\ \mu m$ which is in reasonable agreement with the measured value of $\Lambda_0$ if the high variation for the ionic radius of Ga is taken into account.

References
[S6.1] Shannon *et al.*, Acta Cryst. **B25**, 925-946 (1969).
[S6.2] de Laeter *et al.*, Pure Appl. Chem. **75** (6) 683–800 (2003).

## S7) Thermal penetration depth as a function of the intrinsic phonon mean free path

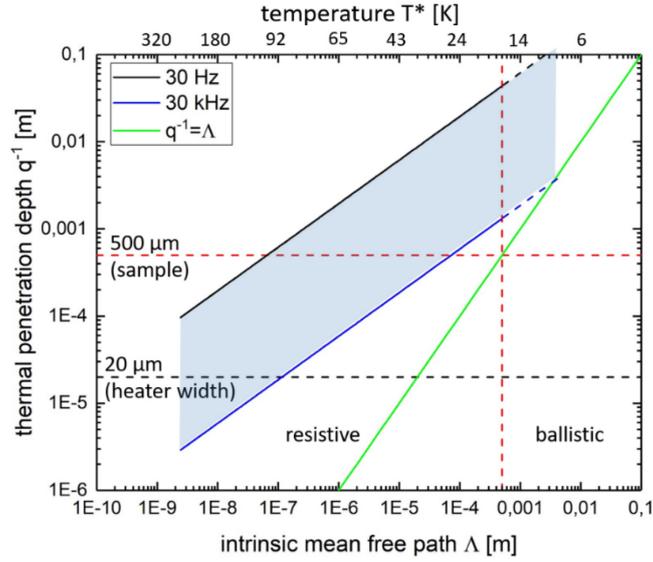

**Fig. S7.** Thermal penetration depth as a function of the intrinsic phonon mean free path (neglecting boundary effects) for various measurement frequencies. For a correct measurement within the resistive transport regime $q^{-1}$ should range between the heater width (20 µm) and the sample thickness (500 µm). Furthermore, it should always be $q^{-1} > \Lambda$. As can be seen, this requires measurements at different frequencies for different temperatures. The frequencies applied in this study raged from 10 Hz to 30 kHz. The vertical line at the sample thickness of 500 µm separates the resistive from the ballistic transport regime. While the relation between thermal penetration depth and intrinsic mean free path is valid for any sample, the temperature scale (denoted with $T^*$) highly depends on sample quality and is only applicable to the samples used in this study.

The linear approximation $\lambda \propto \frac{\Delta \ln 2\omega}{\Delta U_{3\omega}}$ made for evaluation of the measurements is only valid in a specific frequency band. This can be described by the thermal penetration depth $\left|\frac{1}{q}\right| = \sqrt{\frac{D}{2\omega}}$.

This value has to be larger than the heater width while remaining smaller than the overall sample thickness to avoid errors in evaluation. Also, the thermal penetration depth should always exceed the intrinsic phonon mean free path $\Lambda$ to accurately describe all occurring scattering mechanisms. Since the thermal diffusivity $D = \frac{\lambda}{c_p \rho}$ is temperature dependent, this frequency band will differ for every measured temperature. This in principle is depicted in figure S7.

Here, the relation between thermal penetration depth $q^{-1}$ and intrinsic mean free path $\Lambda$ of the phonons is shown for different frequencies within the measurement range. For measurements in the resistive regime, the case of $\Lambda < d$, the condition of $q^{-1} > \Lambda$ has to be met. In the ballistic regime, where $\Lambda > d$ it still has to be $q^{-1} > \Lambda$ and consequently also $q^{-1} >\approx d$.